\documentclass[12pt]{article}
\usepackage[pdftex,colorlinks=true,linkcolor=blue,citecolor=blue,urlcolor=blue]{hyperref}
\usepackage[usenames,dvipsnames]{xcolor}
\usepackage{amsmath,amsfonts,amssymb}
\usepackage{graphicx}
\usepackage{psfrag}
\usepackage{enumerate}
\usepackage{mathrsfs}
\usepackage{dsfont}
\newcommand\blfootnote[1]{%
  \begingroup
  \renewcommand\thefootnote{}\footnote{\hspace{-6mm}#1}%
  \addtocounter{footnote}{-1}%
  \endgroup
}

\def\({\left(}
\def\){\right)}
\def\[{\left[}
\def\]{\right]}

\numberwithin{equation}{section}

\begin{document}
\clearpage\thispagestyle{empty}

\begin{center}


{\large \bf
 A coarse-grained generalized second law for holographic conformal field theories}

%
%

\vspace{7mm}

William Bunting$^{a,b}$, Zicao Fu$^{c,d}$, and Donald Marolf$^{c}$ \\

\blfootnote{\tt wbunting@umd.edu,  zicao\_fu@umail.ucsb.edu, marolf@physics.ucsb.edu}




\bigskip\centerline{$^a$\it Physics Deparment, University of Maryland}
\smallskip\centerline{\it College Park, MD 20742, USA}

\bigskip\centerline{$^b$\it California Institute of Technology}
\smallskip\centerline{\it Pasadena, CA 91125, USA}

\bigskip\centerline{$^c$\it Department of Physics, University of California,}
\smallskip\centerline{\it Santa Barbara, CA 93106, USA}

\bigskip\centerline{$^d$\it Department of Physics, Tsinghua University,}
\smallskip\centerline{\it Beijing, 100084, China}


\end{center}

\vspace{5mm}

\begin{abstract}
We consider the universal sector of a $d >2$ dimensional large-$N$ strongly-interacting holographic CFT on a black hole spacetime background $B$. When our CFT$_d$ is coupled to dynamical Einstein-Hilbert gravity with Newton constant $G_{d}$, the combined system can be shown to satisfy a version of the thermodynamic Generalized Second Law (GSL) at leading order in $G_{d}$. The quantity $S_{CFT} + \frac{A(H_{B, \text{perturbed}})}{4G_{d}}$ is non-decreasing, where $A(H_{B, \text{perturbed}})$ is the (time-dependent) area of the new event horizon in the coupled theory.  Our $S_{CFT}$ is the notion of (coarse-grained) CFT entropy outside the black hole given by causal holographic information -- a quantity in turn defined in the AdS$_{d+1}$ dual by the renormalized area $A_{ren}(H_{\rm bulk})$ of a corresponding bulk causal horizon.   A corollary is that the fine-grained GSL must hold for finite processes taken as a whole, though local decreases of the fine-grained generalized entropy are not obviously forbidden.  Another corollary, given by setting $G_{d} = 0$,
states that no finite process taken as a whole can increase the renormalized free energy $F = E_{out} - T S_{CFT} - \Omega J$, with $T, \Omega$ constants set by ${H}_B$.  This latter corollary constitutes a 2nd law for appropriate non-compact AdS event horizons.
\end{abstract}

\setcounter{footnote}{0}
\newpage
\clearpage
\setcounter{page}{1}

\tableofcontents

%
\section{Introduction}
%
\label{sec:intro}

The thermodynamic properties of classical black holes and the thermal properties of Hawking radiation
are well-established.  Considered together, they are generally taken to indicate that much deeper relations of this sort remain true at the level of quantum gravity.  They may also herald the appearance of new physical principles, yet to be understood.

The conjectured Generalized Second Law (GSL) \cite{Bekenstein:1973ur,Hawking:1974sw} lies
between the classical and fully quantum extremes,  stating
\begin{equation}
\label{GSL}
\frac{d}{d\lambda} S_{total} = \frac{d}{d\lambda} \left(
\frac{1}{4G_d} {\rm Area}_{d-2}(C_{\tilde B}) + S_{QFT}\right) \ge 0,
\end{equation}
where $\tilde B$ is an evolving black hole spacetime of dimension $d \ge 3$ with compactly-generated horizon, $C_{\tilde B}$ is a cut of its horizon, $S_{QFT}$ is the QFT entropy outside the black hole, $G_d$ is Newton's gravitational constant, and $\lambda$ is any parameter that increases toward the future.  One expects that this conjecture can be made precise in perturbative gravity, where one expands the gravitational field of $\tilde B$ around some classical black hole spacetime $B$ in powers of the Planck length or, setting $\hbar =1$, in powers of $G_d$.  For $d = 2$ Einstein gravity becomes trivial, but one can still formulate a GSL in dilaton gravity theories in which the dilaton field plays the role of the area \cite{Fiola:1994ir}.

It is natural to take the classical spacetime $B$ to satisfy the null energy condition, so its horizon area is non-decreasing by the Hawking area theorem \cite{Hawking:1971tu}. If it is increasing, then the $O(G_{d}^{-1})$ contribution to \eqref{GSL} is positive and dominates all perturbative quantum corrections.  The result \eqref{GSL} then follows trivially; new effects can be relevant only when the area of the classical background $B$ is constant.  We therefore assume below that the horizon $H_B$ is generated by a Killing field $\xi_B$  of $B$, and we investigate the first quantum correction to \eqref{GSL}.  This correction contributes to \eqref{GSL} at order zero in $G_d$.

A GSL at this order was established in \cite{Wall:2011hj} for cases where the QFT outside the black hole is super-renormalizeable.  While the result is expected to hold more generally, it is clear that such extensions will require new methods of proof; see e.g. \cite{Bousso:2014uxa}.  In this work we consider the contrasting special case of a large $N$ (and perhaps strongly coupled) holographic conformal field theory (CFT).  I.e., {\it before} coupling the CFT$_d$ to gravity, the theory is dual to a bulk system involving asymptotically (locally) Anti-de Sitter (AlAdS) gravity in $d+1$ spacetime dimensions.  This bulk dual can then be used to study the properties of the CFT.

Our GSL will also differ in that we will use a coarse-grained notion of the entropy $S_{QFT}$ defined by the so-called causal holographic information (\cite{Hubeny:2012wa}, see below), as opposed to the fine-grained (von Neumann) entropy used in \cite{Wall:2011hj}.  In equilibrium this coarse-grained entropy is known to agree with the fine-grained one so, when our system evolves from one equilibrium state to another,  our result implies that the final fine-grained generalized entropy $\frac{1}{4G_d} {\rm Area}_{d-2}(C_{\tilde B}) + S_{QFT}$ must be at least as large as its initial value.  We expect that a stronger result also holds which would forbid even local decreases of the fine-grained generalized entropy at finite intermediate times, though we save its investigation for future work.  Such a fine-grained local GSL should be the most fundamental version, in that one expects it to imply our local coarse-grained GSL through the same argument that leads to the coarse-grained second law of thermodynamics for ordinary closed quantum systems (where unitarity guarantees the fine-grained von Neumann entropy to be independent of time); see e.g. the discussion in \cite{Wall:2013uza}.

Changes in the gravitational entropy $\frac{1}{4G_d} {\rm Area}_{d-2}(C_{\tilde B})$ are easy to analyze at leading order in $G_d$.
If the perturbed system $\tilde B$ settles down at late times to a stationary black hole that is perturbatively close to the original $B$, then at leading  order in $G_d$ the Raychaudhuri equation\footnote{Or the dilaton gravity analogue for $d=2$; see e.g. \cite{Wall:2011kb}.} shows the gravitational entropy on any cut of the horizon to be fully determined by the flux of stress energy across the horizon of $B$.  The calculation is standard, appearing in derivations of the physical process first law such as \cite{Carter,Wald:1995yp,Jacobson:1999mi,Amsel:2007mh}.  At this order one may associate any cut $C_{\tilde B}$ of the back-reacted spacetime $\tilde B$ with a cut $C_B$ of the unperturbed horizon $C_B$. Since caustics do not form in perturbation theory, one merely tracks changes in the area along each generator to find
\begin{equation}
\label{RC}
\frac{1}{4G_d}  \frac{d^2}{d\lambda^2} {\rm Area}_{d-2}(C_{\tilde B}) = - \int_{C_B} \sqrt{\sigma}\,2 \pi T_{\alpha \beta} k^\alpha k^\beta
\end{equation}
in terms of an affine parameter $\lambda$ along $H_B$ (or equivalently at this order, along $H_{\tilde B}$), the affinely parameterized tangents $k^\alpha$ to its generators, and the volume element $\sqrt{\sigma}$ induced on $C_B$. That the relation is 2nd order follows immediately from the 2nd order nature of the Rauchaudhuri equation.

It remains to study the changes in $S_{QFT}$.  Although we study a CFT coupled to gravity, up to contributions from linearized gravitons the leading behavior of $S_{QFT}$ is controlled by properties of the original CFT on the fixed spacetime background $B$.  The former may be addressed separately as in \cite{Wall:2011hj}; we will not mention them again and thus freely replace $S_{QFT}$ by $S_{CFT}$ below.  We will show that changes in this quantity can be written in the form
\begin{equation}
\label{SCFT2part}
\frac{d}{d\lambda} S_{CFT} = \frac{d}{d\lambda}  S_{CFT,\ non-dec} + \frac{d}{d\lambda}  S_{CFT,\ adiabatic},
\end{equation}
where $\frac{d}{d\lambda}  S_{CFT,\ non-dec} \ge 0$,  the term $S_{CFT,\ adiabatic}$ satisfies
\begin{equation}
\label{CFTB}
\frac{d^2}{d\lambda^2} S_{CFT,\ adiabatic}(C_B) = \int_{C_B} \sqrt{\sigma}\,2\pi T_{\alpha \beta} k^\alpha k^\beta,
\end{equation}
and $\frac{d}{d\lambda} S_{CFT,\ adiabatic} \to 0$ in the far future\footnote{Since we have already assumed equilibrium in the far future, this last condition follows trivially for many definitions of $S_{CFT,\ adiabatic}$.}.  The GSL \eqref{GSL} then follows by a straightforward argument.  We simply note that \eqref{CFTB} and \eqref{RC} imply
\begin{equation}
\frac{d^2}{d\lambda^2} \left( S_{CFT,\ adiabatic} + \frac{1}{4G_d}  {\rm Area}_{d-2}(C_{\tilde B}) \right) =0,
\end{equation}
so the combination
\begin{equation}
\frac{d}{d\lambda}  S_{total,\ adiabatic} : = \frac{d}{d\lambda}  \left( S_{CFT,\ adiabatic} + \frac{1}{4G_d}  {\rm Area}_{d-2}(C_{\tilde B}) \right)
\end{equation}
is independent of $\lambda$. But since $\frac{d}{d\lambda} S_{CFT,\ adiabatic}$ and (due to the equilibrium in the far future) $\frac{d}{d\lambda}  {\rm Area}_{d-2}(C_{\tilde B})$ both approach zero, we see that $\frac{d}{d\lambda}  S_{total,\ adiabatic}$ must vanish. Any change in $S_{total}$ is then given by $\frac{d}{d\lambda}  S_{CFT,\ non-dec} \ge 0$, so the GSL holds in the form \eqref{GSL}.

Establishing \eqref{SCFT2part} will be the main focus of our work.  As advertised, it involves only the holographic CFT on the fixed background $B$ -- a context in which the standard AdS/CFT machinery may be applied.  This $B$ then becomes the boundary of an asymptotically (locally) Anti-de Sitter spacetime having bulk Newton constant $G_{d+1}$ that is dual to some particular CFT state.  Since $B$ is a black hole spacetime, the bulk geometry also contains a horizon $H$.  The bulk horizon is {\it non}-compactly generated, with each cross-section extending to meet the conformal boundary at the horizon $H_B$ of $B$.

Below, we use a renormalized version of $\frac{1}{4G_{d+1}}$ times the area of certain cuts $C$ of the bulk horizon $H$ as our coarse-grained entropy $S_{CFT}$ for the CFT in the region $B_{out} \subset B$ outside the horizon $H_B$.  As explained in section \ref{sec:CHI}, one may associate moments of time in $B_{out}$ with so-called causal information surfaces \cite{Hubeny:2012wa} which give the desired cuts $C$ of $H$.  Since the area of such surfaces is bounded below by the area of the extremal surfaces that seem to govern the fine-grained entropy\footnote{I.e., those proposed by Hubeny, Rangamani, and Takayanagi  \cite{Hubeny:2012wa}, in a generalization of the Ryu-Takayanagi proposal \cite{Ryu:2006bv,Ryu:2006ef} for static bulk spacetimes.} \cite{Hubeny:2012wa,Wall:2012uf}, it is natural to expect that $\frac{{\rm Area}(C)}{4G_{d+1}}$ is a coarse-grained measure of CFT entropy; see \cite{Freivogel:2013zta,Kelly:2013aja} for specific proposals of possible corresponding field theory coarse-graining procedures.  We adopt this general point of view for the work below.  Our proof that $S_{CFT}$ satisfies a GSL constitutes quantitative check on this idea.

The present paper will consider only the so-called universal sector of our large $N$ (and perhaps strongly coupled) holographic $d>2$ CFT$_d$, in which the bulk dual is described by pure classical $(d+1)$-dimensional Einstein-Hilbert gravity with a negative cosmological constant.  The main argument is given in section \ref{sec:Proof}, which for simplicity considers only Ricci-flat black hole backgrounds $B$.    Section \ref{sec:notRicciFlat} then extends the result to allow arbitrary metrics on $B$ consistent with the required Killing field, horizon, and the above-mentioned null energy condition.  We postpone discussing the relation to causal holographic information until section \ref{sec:CHI}, which also notes that the argument can be equally-well applied to a class of causal horizons somewhat broader than the simple notion of black hole used in earlier sections.  We close in section \ref{sec:disc} with a discussion of future directions and implications for AlAdS systems with fixed boundary metric $g^{(0)}_{\alpha \beta}$.

\section{Main argument}
\label{sec:Proof}

Consider a holographic CFT on a $d \ge 2$ dimensional spacetime $B$ with a bifurcate Killing horizon and an associated Killing field $\xi_B$.  A bifurcate Killing horizon consists of two components that intersect at the bifurcation surface; for simplicity we choose the future horizon and refer to it as $H_B$ in $B$.  For now, we assume $H_B$ to be the event horizon for some region $B_{out} \subset B$, which we refer to as being outside the horizon. We also require $H_B$ to be compactly generated.  We will comment later in section \ref{sec:CHI} on various extensions.

We take our CFT to be a conformal field theory deformed by relevant operators.  Since the theory is holographic, we may study it using a $(d+1)$-dimensional asymptotically-locally Anti-de Sitter (AlAdS) spacetime with boundary $B$ and appropriate boundary conditions on some set of bulk fields.  While bulk theories dual to holographic CFTs typically also contain extra compact dimensions, Kaluza-Klein reduction of such theories allows one to formally write them in the above form by keeping the entire infinite tower of massive Kaluza-Klein modes.

The region $B_{out}$ of the conformal boundary also defines a bulk event horizon $H$ which is not compactly generated since along each cross-section it must extend to the boundary $B$.  We assume that our system equilibrates at late times in the sense that the bulk approximates a stationary solution outside $H$, with $H$ approaching an associated Killing horizon.  The null energy condition in the bulk means \cite{Gao:2000ga} that the intersection of $H$ with the boundary coincides with the event horizon $H_B$ of $B_{out}$ in $B$; i.e., $H_B = H \cap B$.  As mentioned in the introduction, we will use the area of a cut $C$ of the bulk horizon $H$ as a measure of coarse-grained CFT entropy in $B_{out}$.  We take this $C$ to be sufficiently smooth at $B$ in the conformally compactified AlAdS spacetime so that $C_B = C \cap H_B$ is a cut of $H_B$. Since the second law concerns changes in entropy, we will compare the areas of various cuts $C$ below.

In this section we take the Ricci tensor ${\cal R}_{\alpha \beta}$ to vanish everywhere. Since we consider only the universal sector of our holographic CFT, the bulk dual takes the particularly simple Fefferman-Graham form \cite{FeffermanGraham}

\begin{equation}
\label{GFG}
G_{AB} dx^A dx^B = \frac{\ell^2}{z^2} \[ dz^2 + g_{\alpha \beta}(z)  dx^\alpha dx^\beta \],
\end{equation}
with \cite{deHaro:2000xn}
\begin{equation}
\label{simpleFGA}
g_{\alpha \beta}(z) =  g_{\alpha \beta}^{(0)}  + \frac{16 \pi G_{d+1}}{d \ell^{d-1}} z^d T_{\alpha \beta} +  \dots,
\end{equation}
where $\dots$ denote higher order terms in $z$; see \cite{Fischetti:2012rd} for a review of such expansions addressed to relativists.  It is useful to decompose this metric according to
\begin{equation}
\label{gdecomp}
G_{AB}  = \bar G_{AB} + \delta G_{AB},
\end{equation}
where the background part
\begin{equation}
\label{barG}
\bar G_{AB} dx^A dx^B = \frac{\ell^2}{z^2} \[ dz^2 + \bar g_{\alpha \beta}  dx^\alpha dx^\beta \]
\end{equation}
has
\begin{equation}
\label{barg}
\bar g_{\alpha \beta} = g^{(0)}_{\alpha \beta}
\end{equation}
independent of $z$,
and the deviation
\begin{equation}
\label{dGab}
\delta G_{AB} dx^A dx^B = \frac{\ell^2}{z^2} \delta g_{\alpha \beta}  dx^\alpha dx^\beta,
\end{equation}
has
\begin{equation}
\label{dgab}
\delta g_{\alpha \beta} = \frac{16 \pi G_{d+1}}{d \ell^{d-1}} z^d T_{\alpha \beta} +  \dots = O(z^d).
\end{equation}
We will also use the notation $g_{AB} dx^A dx^B :=  g_{\alpha \beta} (z) dx^\alpha dx^\beta$ and similarly for other tensors with boundary indices.

Since $\bar G_{AB}$ is fully determined by boundary conditions at $B$, from the bulk perspective it defines a non-dynamical background about which we may perturb.  We will use the decomposition \eqref{gdecomp} only near the $z=0$ boundary, so possible singularities at $z= \infty$ in $\bar G_{AB}$  are of no concern. Note that the Killing field $\xi_B$ of $g^{(0)}_{\alpha \beta}$ implies a corresponding Killing field $\bar \xi = \xi_B^\alpha \partial_\alpha$ of $\bar G_{AB}$.  The Killing horizon $\bar H$ of $\bar \xi$ may be defined by the equation
\begin{equation}
\label{Hproj}
x^\alpha = x^{\alpha}_B
\end{equation}
where $x^{\alpha}_B$ ranges over points in $H_B$.

We wish to study cuts $C$ of the full event horizon $H$ for which $C \cap B$ is a cut $C_B$ of $H_B$.  A first step is to show that they admit a useful notion of renormalized area.  Since $\delta g_{\alpha \beta} = O(z^d)$, it follows from \eqref{Hproj} that points $p$ in $C$ satisfy
\begin{equation}
\label{Cproj}
x^\alpha = x^{\alpha}_B(p) + O(z^d)
\end{equation}
for some smooth map $x^{\alpha}_B$ from $C$ to $C_B$.
In particular, the action of any such map on the constant $z$ slices of $C$ defines a family of cuts $C_B(z)$ of $H_B$.  And since $H_B$ is a Killing horizon, all cuts $C_B(z)$ have the same area as $H_B$ itself.  As a result, we have
\begin{equation}
{\rm Area}_{d-1} (C) = {\rm Area}_{d-2}(H_B) \int \frac{\ell^{d-1} dz}{z^{d-1}} \left(1 + O(z^d)\right),
\end{equation}
where the $O(z^d)$ corrections come both from $\delta g_{AB}$ and the error term in \eqref{Cproj}.\footnote{Displacements $\delta x = x^\alpha - x^\alpha_B(p)$ along the cut just reparametrize the surface and can be ignored.  Changes in the induced metric are then quadratic in the remaining normal displacements, so contributions from the error term in \eqref{Cproj} are in fact suppressed by $z^{2d}$.}  We may thus define a finite renormalized area by introducing the restrictions $C_{z>z_0}$ of $C$ to the region $z>z_0$ and writing
\begin{equation}
\label{Aren}
{\rm Area}_{ren} (C) :=  \lim_{z_0 \rightarrow 0} \left( {\rm Area}_{d-1}(C_{z>z_0}) - \frac{1}{d-2} \frac{\ell^{d-2}}{z_0^{d-2}}{\rm Area}_{d-2}(H_B) \right).
\end{equation}
Note that this is the same expression that would be used to renormalize the area of an extremal surface in this spacetime; see e.g. \cite{Ryu:2006bv,Ryu:2006ef}.  In fact, the area of $C$ differs by only a finite amount from the area of an extremal surface anchored on $C_B$.  We also note that the counterterm in \eqref{Aren} is fully determined by the boundary conditions and is completely independent of the choice of $C$.  So for fixed boundary conditions we may treat all changes in the (unrenormalized) area of $C$ as being finite as well.  Here it is critical that $H_B$ is a Killing horizon (in some conformal frame).  The general behavior of area-divergences for bulk causal horizons is much more complicated \cite{Freivogel:2013zta}.

As noted in section \ref{sec:intro}, our GSL will follow if we can establish \eqref{SCFT2part}. Using our coarse-grained entropy we have $S_{CFT}(C_B) : = \frac{{\rm Area}_{ren}(C)}{4G_d}$ for a certain cuts $C$ intersecting the boundary on $C_B.$  The derivative
$\frac{d}{d\lambda} {\rm Area}_{ren}(C) = \frac{d}{d\lambda} {\rm Area_{d-1}}(C)$ can then be divided into two parts.  The first is the rate at which area is created in the bulk as determined by both the local divergence of tangents to the generators of $H$ and the rate at which generators are added to the horizon.  Since we assume the CFT to reach equilibrium in the far future, the bulk must settle down to a stationary black hole.  That the bulk area creation term is non-negative is then just the usual Hawking area theorem \cite{Hawking:1971tu}.  As a result, after multiplying by $\frac{1}{4G_{d+1}}$ we take this term to be the  $\frac{d}{d\lambda}S_{CFT,\ non-dec}$ of \eqref{SCFT2part}.

\begin{figure}[t]
\centerline{
\includegraphics[width=0.4\textwidth]{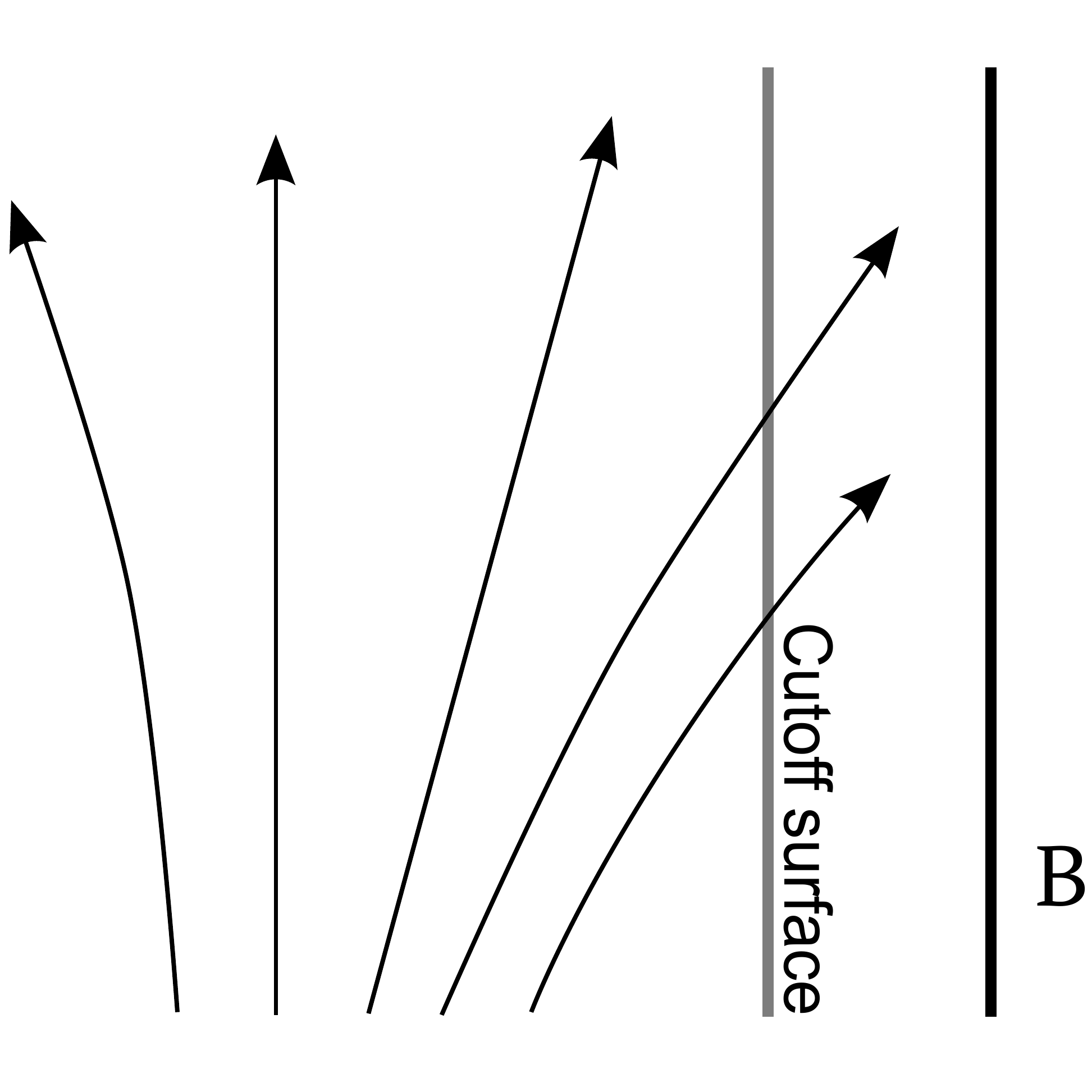}
}
\caption{Horizon generators can flow through our cutoff surface $z=z_0$.
If this behavior persists at arbitrarily small $z_0$, it constitutes a flow through the boundary and there is an associated change of the renormalized coarse-grained entropy of our CFT$_d$ in $B_{out}$. Our final result shows that this flux is directed outward so that
${\rm Flux}(C)$ as defined in the main text is negative at finite times.}
\label{fig:flow}
\end{figure}

The remaining contribution to $\frac{d}{d\lambda} {\rm Area_{d-1}}(C)$ is then the rate ${\rm Flux}(C)$ at which area flows inward through the boundary $B$ at $C_B$; see figure \ref{fig:flow}. To identify $\frac{{\rm Flux}(C)}{4G_{d+1}}$ with $\frac{d}{d\lambda} S_{CFT,\ adiabatic}$ we must show
\begin{equation}
\label{CFTB2}
\frac{d}{d\lambda} \frac{{\rm Flux}(C)}{4G_{d+1}}  = \int_{C_B} \sqrt{\sigma} 2 \pi T_{\alpha \beta} k^\alpha k^\beta,
\end{equation}
where we remind the reader that $k^\alpha$ is an affinely parametrized generator of $H_B$.
We proceed by computing the flux though surfaces of constant $z=z_0$ and taking the limit $z_0 \rightarrow 0$.  We will make use of the tangents $U^A = \frac{d}{d\lambda} x^A$, $\bar U^A = \frac{d}{d\lambda} x^A$ to the generators of $H$ and $\bar H$ and the (inward pointing) unit normal one-form $n = \frac{\ell}{z}dz$ to the cut-off surfaces $z = z_0$.  We choose the affine parameter on $\bar H$ so that $\bar U = k^\alpha \partial_\alpha$, and we take that on $H$ to differ only by an amount first order in $\delta G_{AB}$.

Note that
\begin{equation}
\label{FluxC}
{\rm Flux}(C) = \lim_{z_0 \rightarrow 0} \int_{C|_{z= z_0}} \sqrt{\omega_{z_0}}\, U^A n_A,
\end{equation}
where $C|_{z= z_0}$ is the intersection of $C$ with the surface $z=z_0$ and $\sqrt{\omega_{z_0}}$ is the induced volume element on this surface.  The volume element $\sqrt{\omega_{z_0}}$ diverges as $z_0 \rightarrow 0$, but we will show that $U^An_A$ vanishes fast enough that ${\rm Flux}(C)$ is finite. Since our goal is to establish \eqref{CFTB2} to leading order,
and since $\sqrt{\omega_{z_0}}$ is independent of $\lambda$, we may in fact study $\frac{d}{d\lambda} U^An_A$. It is sufficient to show
\begin{equation}
\label{CFTB3}
\frac{d}{d\lambda} (n_A U^A)  = \left(\frac{z}{\ell} \right)^{d-2} 8 \pi  G_{d+1} T_{\alpha \beta} k^\alpha k^\beta + \dots,
\end{equation}
where the $\dots$ represent subleading terms.
If desired, the actual value of $U^An_A$ may then be obtained by integrating over $\lambda$ using the assumption of equilibrium in the far future (i.e., $U^An_A \rightarrow 0$ in the far future at each $z$).

We will see below that finite contributions to ${\rm Flux}(C)$  are first order in $\delta G_{AB}$; all higher contributions vanish as $z_0 \rightarrow 0$.  Since $\bar U^An_A = 0$, we may thus neglect contributions to ${\rm Flux}(C)$ associated with differences in either the locations or the affine parameters of $H$ and $\bar H$.  We may then replace $\frac{d}{d\lambda} = U^A \nabla_A$ by $\bar U^A \bar \nabla_A$, where $\nabla_A, \bar \nabla_A$ are the covariant derivatives defined respectively by $G_{AB}$ and $\bar G_{AB}$. In a slight abuse of notation, we simply write
$\frac{d}{d\lambda} = U^A \nabla_A = \bar U^A \bar \nabla_A$ below. Since the one-form $n = \frac{\ell}{z}dz$ is independent of $\delta G_{AB}$, to investigate $\frac{d}{d\lambda} (n_A U^A)$ we need only define $\delta U^A = U^A - \bar U^A$ and compute \begin{equation}
\label{1stordercalc}
\frac{d}{d\lambda} (n_A U^A) = \bar U^B \bar \nabla_B (n_A \delta U^A) = \delta U^A \bar U^B \bar \nabla_B n_A + n_A \bar U^B \bar \nabla_B \delta U^A.
\end{equation}

Let us begin with the first term on the right-hand side.  Introducing the Christoffel symbols $\bar \Gamma^C_{AB}$ of \eqref{barG}, we find
\begin{equation}
\label{gradn}
\bar U^B \bar \nabla_B n_A = \frac{\ell}{z} \bar U^B \bar \Gamma^z_{AB} = - \ell z^{-2} \bar U^B g_{BA}^{(0)} = -\ell^{-1} \bar U^B \bar G_{AB}.
\end{equation}
Thus
\begin{equation}
\label{partof1}
\delta U^A \bar U^B \bar \nabla_B n_A =  -\ell^{-1} \delta U^A  \bar U^B \bar G_{AB}.
\end{equation}
But $U = \bar U + \delta U$ is null with respect to $G_{AB}$, so to linear order we have
\begin{equation}
\label{norm}
0 = \delta (U^2) = 2 \bar U^B \bar G_{AB} \delta U^A + \bar U^A \bar U^B \delta G_{AB},
\end{equation}
which yields
\begin{equation}
\label{term1}
\delta U^A \bar U^B \bar \nabla_B n_A = \frac{1}{2\ell} \bar U^A \bar U^B \delta G_{AB} = \left( \frac{z}{\ell}\right)^{d-2} \frac{8 \pi G_{d+1}}{d} T_{\alpha \beta} \bar U^\alpha \bar U^\beta +\dots \ .
\end{equation}
In particular, since $\sqrt{\omega_z} =  (\ell/z)^{(d-2)} \sqrt{\sigma} + \dots$, inserting \eqref{term1} into \eqref{FluxC} yields a finite result as desired.  Since $d>2$, \eqref{term1} vanishes as $z \to 0$, so higher order contributions are smaller and can be neglected.  It is even of the form \eqref{CFTB3}, though with a different coefficient.

To compute the second term on the right-hand side of \eqref{1stordercalc}, we study the first-order effect of $\delta G_{AB}$ on the geodesic equation $U^A \nabla_A U^B =0$.  Contracting the first variation of this equation with $n_B$ gives
\begin{equation}
\label{varygeodesic}
0 = n_B \left(\delta U^A \bar \nabla_A \bar U^B + \bar U^A \delta \Gamma_{AC}^B \bar U^C + \bar U^A \bar \nabla_A \delta U^B  \right),
\end{equation}
where $\delta \Gamma_{AC}^B = \Gamma_{AC}^B  - \bar \Gamma_{AC}^B$, $\Gamma_{AC}^B$ are the Christoffel symbols of $G_{AB}$, and we include only first order terms in $\delta G_{AB}$.
The term we need to understand to finish the computation of \eqref{1stordercalc} is the final one in \eqref{varygeodesic}, and so may be found by analyzing the first two terms in this equation. The identity
\begin{equation}
\label{2isminus1}
\bar \nabla _B (n_A \bar U^A) = \bar U^A \bar \nabla _B n_A + n_A \bar \nabla _B \bar U^A
\end{equation}
shows that the first term in \eqref{varygeodesic} is again \eqref{term1} up to a factor of $-1$. And the second term is readily computed using
\begin{equation}
\label{deltaGamma}
n_A \delta \Gamma_{AC}^B = - \frac{d-2}{2} \ell^{d-1} z^{d-2} \frac{ 16 \pi  G_{d+1}}{d} T_{\alpha \beta} + \dots ,
\end{equation}
which follows immediately from \eqref{dGab} and \eqref{dgab}. Adding up all of the contributions then verifies \eqref{CFTB3} as desired.  This completes the proof of \eqref{GSL} for the simple case considered here.

\section{Beyond Ricci flatness}
\label{sec:notRicciFlat}

The argument above was straightforward, but applied only to perturbations of Ricci-flat stationary spacetimes $B$. We now drop the Ricci-flat assumption, though we continue to focus on the universal sector of our holographic CFT$_d$.  As a result, the bulk theory in this section is again just pure $(d+1)$-dimensional Einstein-Hilbert gravity with cosmological constant; no matter fields are allowed.  In Fefferman-Graham coordinates, the bulk metric $G_{AB}$ remains of the form \eqref{GFG}, though now with the expansion \cite{deHaro:2000xn}

\begin{equation}
\label{FGAR}
g_{\alpha \beta}(z) =  g_{\alpha \beta}^{(0)}  + z^2 g_{\alpha \beta}^{(2)} + \cdots + z^{2n} g_{\alpha \beta}^{(2n)} + z^d \bar g_{\alpha \beta}^{(d)} + z^d \log{z^2} \; g_{\alpha \beta}^{(d)} + \frac{16 \pi G_{d+1}}{d \ell^{d-1}} z^d T_{\alpha \beta}     + \dots \ .
\end{equation}
The $g_{\alpha \beta}^{(n)}$ for $n > 0$ are rank 2 symmetric tensors with scaling dimension $n$ (counting powers of inverse length) constructed locally from $g_{\alpha \beta}^{(0)}$, and
 $\bar g_{\alpha \beta}^{(d)}$  is a similar tensor of scaling dimension $d$.   As an example, for $d > 2$  we have
\begin{equation}
g_{\alpha \beta}^{(2)} = - \frac{1}{d-2} \left({\cal R}_{\alpha \beta} - \frac{1}{2(d-1)} {\cal R} g^{(0)}_{\alpha \beta} \right)
\end{equation}
where ${\cal R}_{\alpha \beta}, {\cal R}$ are respectively the Ricci tensor and scalar of $g^{(0)}_{\alpha \beta}$.

We wish to generalize the argument of section \ref{sec:Proof} to deal with the additional terms $\bar g_{\alpha \beta}^{(d)}$
 and $g_{\alpha \beta}^{(2n)}$ for $2n <d$.  Since these terms are fully determined by the metric $g^{(0)}_{\alpha \beta}$ on $B$, it is natural to include them in the background metric $\bar G_{AB}$, replacing \eqref{barg} with the finite sum
 \begin{equation}
 \label{barg2}
 \bar g_{\alpha \beta}(z) = \left(
 \sum_{{n \in \mathbb{Z}^+ \cup \{0\} } \atop {n \le d/2} } z^{2n} g_{\alpha \beta}^{(2n)}
 \right)
 + z^d \log{z^2} \; \tilde g_{\alpha \beta}^{(d)}   .
 \end{equation}
 As a result, we again have the decomposition \eqref{gdecomp} with $\delta G_{AB}$ given by \eqref{dGab} and \eqref{dgab} and where $\bar G_{AB}$ again has an exact symmetry generated by the Killing field $\bar \xi = k^\alpha \partial_\alpha$.

A key observation is that the corresponding bulk horizon remains a Killing horizon generated by $\bar \xi$.  To see this, define $\bar H$ to be the lift of $H_B$ to the bulk via \eqref{Hproj}.  It suffices to show that $\bar H$ is a null surface with null tangent $\bar \xi$; i.e., that every vector tangent to $\bar H$ is orthogonal to $\bar \xi$. This ensures that $\bar H$ is a Killing horizon, and in fact the bulk event horizon defined by $\bar G_{AB}$.

We argue as follows.  Any vector $v$ tangent to $\bar H$ is of the form $v = a \partial_z + v_B^\alpha \partial_\alpha$ where $v^B$ is tangent to $H_B$.  It is clear from \eqref{GFG} that $\partial_z$ is orthogonal to $\bar \xi = k^\alpha \partial_\alpha$.  So it remains to show only that
\begin{equation}
\label{Hbarnull}
v_B^\alpha \bar g_{\alpha \beta} k^\beta =0 \ \ \ {\rm on} \ \ \ \bar H,
\end{equation}
or equivalently that   $v_B^\alpha  g^{(n)}_{\alpha \beta} k^\beta =0$ on $\bar H$ for all $n$ (and correspondingly for $\tilde g^{(d)}$). To begin, consider the case $v_B^\alpha \propto k^\alpha$ and note that  $ \xi_B^\alpha g^{(n)}_{\alpha \beta} \xi_B^\beta$ is invariant under the isometry generated by $\xi_B$ and thus constant over each generator of $\bar H$ to either the future or past of the bifurcation surface.  It also vanishes at the bifurcation surface itself, where $\xi_B =0$.  Since our quantity is continuous, it must thus vanish on all of $\bar H$.  So $k^\alpha g^{(n)}_{\alpha \beta} k^\beta = \frac{1}{\lambda^2} \xi_B^\alpha g^{(n)}_{\alpha \beta} \xi_B^\beta$ vanishes on $\bar H$ as well, where we have again used continuity to reach this conclusion at the bifurcation surface.

The argument for the remaining components of \eqref{Hbarnull} can be stated in an identical form.  To do so, we complete a basis for the tangent space at each point in $\bar H$ using vector fields Lie-dragged by $k = \frac{d}{d\lambda}$ along each generator from the bifurcation surface.  Since
$B$ satisfies the null energy condition, $H_B$ satisfies the zeroeth law of horizon mechanics (see e.g. \cite{Wald:1984rg}) and has constant surface gravity $\kappa$.  This allows us to write $\xi_B = \kappa \lambda \frac{d}{d\lambda}$, showing that the above vector fields are Lie-derived by $\xi_B$ as well $k$ and so define the desired invariants.  We conclude that $\bar H$ is a null surface with generators proportional to $\bar \xi$, and thusa Killing horizon as claimed above.  In particular, $\bar U = k^\alpha \partial_\alpha$ again gives an affinely-parametrized set of tangents to these generators. We expect a similar result to hold even for non-bifurcate horizons; see e.g. \cite{Reall:2014pwa} for a discussion of terms that do not involve derivatives of the Riemann tensor.

We may now follow the general outline of the argument in section \ref{sec:Proof}.  As before, we begin by discussing possible divergences associated with the area of our cut $C$.  Contributions from $\delta G_{AB}$ are finite, so we focus on $\bar G_{AB}$.   The new terms in $\bar G_{AB}$ at order $d-2$ and below contribute additional divergences to \eqref{Aren}, but these may be canceled by the standard counter-terms (locally constructed from $g^{(0)}_{\alpha \beta}$ alone) used to renormalize the area of an extremal surface.  Indeed, the Killing symmetry of our background requires these counter-terms to take the same values for all cuts $C$, including the case where $C$ is the extremal surface that forms the bifurcation surface of $H_B$.  So for fixed boundary conditions changes in ${\rm Area}_{d-1}(C)$ are again finite and coincide with changes in ${\rm Area}_{ren}(C)$.

As in section \ref{sec:Proof}, we now rely on the bulk Hawking area theorem to identify $\frac{d}{d\lambda} S_{CFT,\ non-dec}$ with the rate at which bulk area is produced.  It then remains to generalize the argument for \eqref{CFTB3}.  Equations \eqref{1stordercalc}, \eqref{norm}, \eqref{varygeodesic}, \eqref{deltaGamma}, and \eqref{2isminus1} are all unchanged, so we need only consider contributions to \eqref{term1} that come from corrections to \eqref{gradn} (and thus \eqref{partof1}) or from new terms in $\delta G_{AB}$ in \eqref{norm}.

These corrections are all proportional to
\begin{equation}
\label{newgterm}
z^{n-2} g^{(n)}_{AB} \bar U^B \delta U^A,
\end{equation}
and similarly for $\bar g^{(d)}$.  By \eqref{Hbarnull},  the terms \eqref{newgterm} depend only on the part of $\delta U^A$ both transverse to $H_B$ and parallel to some constant $z$ surface, which we characterize by $\delta U^k := \left( \delta U^\alpha k^\beta g_{\alpha \beta}^{(0)} \right) $ .  Indeed, introducing a future-pointing null vector field $l^\alpha$ on $B$ that satisfies $l^\alpha k^\beta g_{\alpha \beta}^{(0)} = -1$ one may write
\begin{equation}
\delta U^A \partial_A= \delta U^z \partial_z + \delta U^\alpha_{H_B}\partial_\alpha - \delta U^k \ l^\alpha \partial_\alpha,
\end{equation}
where $\delta U_{H_B}^\alpha$ is tangent to $H_B$ at each $z$. Our previous results \eqref{norm} and \eqref{term1} then require $\delta U^k = O(z^d)$ so that \eqref{newgterm} vanishes at least as $z^{d+n-2}$.  Since $n \ge 2$, such behavior is subleading relative to the $O(z^{d-2})$ terms displayed in \eqref{term1}  and \eqref{CFTB3} and so gives no contribution to \eqref{FluxC}.  We may thus continue to identify $\frac{{\rm Flux}(C)}{4G_{d+1}}$ with $\frac{d}{d\lambda} S_{CFT,\ adiabatic}$ and conclude that the GSL \eqref{GSL} holds as desired.

\section{Relation to Causal Holographic Information}
\label{sec:CHI}

We now explain the relation of our construction to causal holographic information \cite{Hubeny:2012wa}.  Along the way, we explain the sense in which our result can be generalized to an arbitrary Killing horizon $H_B$ of $B$.

As noted above, we consider an asymptotically locally AdS$_{d+1}$ bulk spacetime $M$ with a choice of conformal frame for which the boundary spacetime $B$ contains a compactly-generated future Killing horizon $H_B$; this means that the space of null generators for $H_B$ is a compact manifold without boundary.   We require $B$ to be connected, and also suppose that there is a closed connected region $B_{out} \subset B$ for which

\begin{enumerate}

\item{} $B_{out}$ contains the past of $H_B$ in $B$.

\item{} $B_{out}$ is invariant under diffeomorphisms generated by the Killing field $\xi_B$ that generates $H_B$.

\item{} $H_B$ lies in the boundary of the past of $B_{out}$.

\end{enumerate}
We will call  $B_{out}$ the outside of $H_B$; since it is closed, this set contains  $H_B$.
Below, we consider cross sections $C_B$ of $H_B$ given by compact submanifolds (without boundary) of $H_B$ which intersect each generator once.


In general, the boundary of the past of $B_{out}$ in $B$ may be disconnected.  In this case $H_B$ need not contain all such components.
For example, when $B$ is a conformal compactification of the asymptotically flat Schwarzschild black hole and $B_{out}$, we may take $H_B$ to be the usual black hole horizon without including null infinity.  Motivated by this example, we use the notation $I^+(B_{out})$ to denote the collection of connected components not in $H_B$; i.e., $H_B \cap I^+(B_{out}) = \emptyset$ and $H_B \cup I^+(B_{out})$ denotes the full boundary of the past in $B$.  This $I^+(B_{out})$ may be null (as when the CFT lives on an asymptotically flat black hole spacetime), timelike, spacelike, of mixed type, or empty.

The region $B_{out}$ also defines a bulk event horizon $H \subset M$.  We take $H$ to include the full boundary of the past of $B_{out}$ lying in the interior of $M$.  Since it will be convenient to think of $H$ as intersecting the boundary, we in fact define $H$ to be the closure of the above set in the given conformal compactification.  Then $H \cap B = H_B \cup I^+(B_{out}).$ Recall that, since the bulk metric is dynamical, this bulk $H$ will not generally be a Killing horizon.

We wish to interpret the areas of certain cross-sections of $H$ as computing a time-dependent coarse-grained entropy in the dual CFT.
To explain this further, recall that a covariant notion of ``moment of time'' in
$B_{out}$ is naturally defined by a future-Cauchy surface $\Sigma_B$ of $B_{out}$.  By this we mean an closed achronal surface for which all points of $B_{out}$ to its future lie in its domain of dependence (aka causal development) $D(\Sigma_B)$ within $B_{out}$.  Note that for later use we take $D(\Sigma_B) \subset B_{out}$ by definition, though for the present purpose we could just as well have discussed the domain of dependence of $\Sigma_B$ within the entire boundary spacetime $B$.  We will also require that all points of $B_{out}$ lie either to the future, to the past, or on the surface $\Sigma_B$.  We will then  say that $\Sigma_B$ is a complete moment of time in $B_{out}$.

Any boundaries of $\Sigma_B$ must lie in $H_B \cup I^+(B_{out})$. In particular, the intersection $\Sigma_B \cap H_B$ is a cross-section $C_B$ of $H$.  We consider time evolutions that keep fixed any intersection of $\partial \Sigma$ with $I^+(B_{out})$; we should take $H_B$ to include any components along which $\partial \Sigma$ is to be moved\footnote{Cases like the asymptotically flat black hole can be included in our treatment even when we wish to move the slice forward along the corresponding null infinity.  Although this null infinity is only a conformal Killing horizon of $B$ --  and not a strict Killing horizon -- in any conformal frame where it forms a smooth part of the AlAdS boundary, the corresponding bulk must nevertheless have an approximate Killing symmetry that can be used just as in section \ref{sec:Proof}.}.

The moment of time $\Sigma_B$ in $B_{out}$ can also be used to specify a cross-section $C$ of the bulk horizon $H$ by constructing the so-called causal information surface \cite{Hubeny:2012wa}\footnote{We thank Aron Wall for suggesting this connection.}.  Consider the above domain of dependence $D(\Sigma_B)$ (defined by inclusion of $\Sigma_B$ in $B_{out}$), and the past and future causal horizons defined by $D(\Sigma_B)$ in the bulk spacetime $M$.  Let $C$ be the intersection of these horizons.  Since $\Sigma_B$ is a complete moment of time in $B_{out}$, the past of $D(\Sigma_B)$ coincides with the past of $B_{out}$ and this bulk future horizon coincides with $H$.  Thus $C \subset H$.  We leave open the question of whether $C$ is in any sense a complete cross section, though it will tend to be so in typical applications.  This $C$ is precisely the causal information surface of \cite{Hubeny:2012wa} associated with $\Sigma_B$ by taking $B_{out}$ to be the boundary spacetime.\footnote{In principle, this might differ from the causal information surface defined by taking the boundary spacetime to be $B$.  But since $B_{out}$ is invariant under $\xi_\partial$, we can always choose a conformal frame in which the boundary spacetime is naturally just $B_{out}$ and in which $\xi_\partial$ remains a Killing field of the boundary spacetime.}

Recall \cite{Hubeny:2012wa,Wall:2011hj} that the area of $C$ is at least as large as that of the extremal surface identified  as computing the von Neumann entropy of the CFT in $D(\Sigma_B)$ by the HRT proposal \cite{Hubeny:2007xt}.\footnote{This is the smallest (real) extremal surface homologous to $\Sigma_B$ and having boundary $\partial \Sigma_B$.} It is thus natural to conjecture $S_{CFT} : = \frac{{\rm Area}(C)}{4G_{d+1}}$ to be some coarse-grained measure of CFT entropy at the moment of time $\Sigma_B$; see \cite{Freivogel:2013zta,Kelly:2013aja} for specific proposals.     The fact that $C$ depends on the choice of $B_{out}$ as well as $\Sigma_B$ indicates that the coarse-graining is specified by the full $D(\Sigma_B)$ and is not specified locally at $\Sigma_B$; this is explicitly the case for the proposal of \cite{Kelly:2013aja}.

\section{Discussion}
\label{sec:disc}

We have shown for $d >2$ that coupling the universal sector of a holographic CFT$_d$ to dynamical gravity leads to a coarse-grained thermodynamic generalized second law \eqref{GSL} at first order in $G_d$ about a background satisfying the null energy condition and having a bifurcate Killing horizon\footnote{Our results also apply to linearized fluctuations about the Hartle-Hawking state for $d=2$.}.  The coarse-grained entropy of our holographic CFT$_d$ was defined using the causal holographic information (CHI) of \cite{Hubeny:2007xt}.  Since in equilibrium CHI coincides with holographic HRT (and Ryu-Takayanagi) entropy, our result implies that for processes both beginning and ending in equilibrium the change in fine-grained generalized entropy defined by HRT/RT entropy is non-negative as well.   But studying the validity of a local GSL for fine-grained HRT entropy remains a task for future work.

An interesting point is that, in our context, the divergences of CHI agree precisely with those of the HRT entropy.  So they can both be renormalized by including the same counter-terms, and the difference between them is finite.
This is related to the well-known fact that causal holographic information and HRT coincide at the bifurcation surface of a bulk Killing horizon.
We have simply identified a useful generalization to arbitrary bulk solutions for which the conformal boundary $B$ admits a (conformal) Killing horizon $H_B$, so that the asymptotic behavior of both HRT and CHI surfaces near the AlAdS boundary is the same as in the case of an exact bulk symmetry.   This is in sharp contrast with the general situation, in which CHI is infinitely larger than the fine-grained HRT entropy \cite{Hubeny:2007xt,Hubeny:2012wa} and, moroever, the divergence is non-local \cite{Freivogel:2013zta}.

Phrased in the form \eqref{GSL}, the GSL concerns the effects of coupling a CFT to dynamical gravity.  But \eqref{SCFT2part} and \eqref{CFTB} involve only the field theory in a fixed black hole background.  So one expects the GSL to have implications even for the fixed-background CFT obtained in the limit $G_d \rightarrow 0$. The entropy of our $d$-dimensional black hole diverges in this limit, and for many purposes it then simply provides an infinite heat bath for the CFT.  In standard thermodynamic settings, applying the second law to a system interacting with a large heat bath tells us that the free energy will not decrease.  A corresponding result can be derived here when the stress-energy flux $T_{\alpha \beta} k^\alpha k^\beta$ on $H_B$ vanishes sufficiently rapidly in the far past and future.  So long as the initial and final states of the black hole are perturbatively close, the so-called physical process first law states that the total change in the gravitational entropy term of \eqref{GSL} is the black-hole temperature $T$ times the total change $\Delta E_{\tilde B}$ in the energy of the black hole up to the usual work term $\Omega J$ associated with the black hole's angular velocity $\Omega$.  (The universal sector studied here carries no charge, so no charge transport term $\Phi \Delta Q$ can appear.) Conservation of energy and \eqref{GSL} then require
\begin{equation}
\label{nonintF}
0 \ge \Delta F =  \Delta E - \Omega \Delta J -  T \Delta S_{CFT}.
\end{equation}
This result can also be derived directly from \eqref{SCFT2part}, \eqref{CFTB} by following the steps in the argument for the physical-process first law.  In particular, the $\Omega \Delta J$ term arises as usual from the difference between the horizon-generating and time-translation Killing fields.   In the form \eqref{nonintF}, our result is a second law for the open system defined by the holographic CFT interacting with a heat bath, or equivalently by the non-compact horizon in the AlAdS bulk.  In particular, as suggested in \cite{Marolf:2013ioa} it constrains the behavior of the so-called black funnels and black droplets reviewed in that reference.

It is interesting to generalize our argument so that it also applies to the
solutions called {\it detuned} droplets and funnels in \cite{Fischetti:2012vt,Marolf:2013ioa} (also discussed earlier in \cite{Marolf:2010tg}).  Simple examples of such solutions turn out to be given by the hyperbolic black holes of \cite{Emparan:1998he,Birmingham:1998nr,Emparan:1999gf}, which were called topological black holes in those references.  These smooth bulk spacetimes admit a conformal boundary $B$ with smooth metric $g^{(0)}_{\alpha \beta}$ having a Killing horizon $H_B$ with surface gravity $\kappa_B$.  There is also a corresponding bulk Killing horizon $\bar H$ having its own surface gravity $\kappa$.  The parameters $\kappa$ and $\kappa_B$ can be chosen independently, so that generically $\kappa \neq \kappa_B$.   This discrepancy is the detuning mentioned above.   In a dual CFT, such solutions describe a state of the field theory at temperature $T$ on a black hole background with surface gravity $\kappa_B \neq 2 \pi T$.  Such states are singular on $H_B$ and, as a result, the conformally compactified bulk fails to be smooth at the conformal boundary.  While it admits a Fefferman-Graham expansion of the form \eqref{FGAR} at orders $z^n$ with $n < d$, and while the expansion continues to all orders away from $H_B$, the order $z^d$ term that would define the boundary stress tensor diverges at $H_B$.

Bulk solutions that approach those discussed above near $H_B$ describe more general states of the dual CFT subject to similar thermal boundary detuned conditions at $H_B$.  For such cases, despite the singularity at $H_B$, the essential elements of our arguments from sections \ref{sec:Proof} and \ref{sec:notRicciFlat} nevertheless go through.   One simply uses the relevant unperturbed hyperbolic black hole bulk solution as the background $\bar G_{AB}$.  The boundary stress tensor $\bar T_{\alpha \beta}$ for this background is now non-zero and in fact divergent on the horizon, but has $\bar T_{\alpha \beta} k^\alpha k^\beta =0$.  So as long as the  contributions $\delta T_{\alpha \beta}$ of $\delta G_{AB}$ to the boundary stress tensor are finite, taking the limit as one approaches $H_B$ from $B_{out}$ again yields \eqref{CFTB3}. While it makes no sense to couple such singular holographic CFT states to dynamical gravity, it nevertheless follows  as in \eqref{nonintF} that equilibrium to equilibrium processes cannot increase the free energy of our system.  Note that the relevant temperature in \eqref{nonintF} is defined by the surface gravity of the bulk horizon of $\bar G_{AB}$, so $T = \frac{\kappa}{2\pi} \neq \frac{\kappa_B}{2\pi}$.

There are also many possible further extensions of our results.
Using \cite{Wall:2015raa} one may readily include perturbative contributions from bulk higher derivative terms  -- corresponding either to $1/N$ effects or, in some cases, corrections to the strong coupling limit -- as well as the case where the holographic CFT$_d$ to couples to higher derivative gravity.  We also expect to be able to allow non-compactly generated horizons  $H_B$,  or cases where $B$ itself has a timelike boundary.  In these situations a GSL should follow from suitable choices of boundary conditions, in the latter case using the approach of either \cite{Aharony:2010ay} or \cite{Takayanagi:2011zk}.

One should also discuss more general CFTs.  Even restricting to those that can be studied holographically, we have thus far considered only the universal sector.  But a forthcoming work \cite{MoreGSL} will move beyond this to address general (large $N$ strongly coupled) holographic theories.  This analysis will in particular include effects from new dynamical divergences in the HRT entropy which depend on expectation values of dynamical operators and thus are not fully determined by sources.   Despite previous claims to the contrary \cite{Hung:2011ta}, it turns out that such dynamical divergences arise generically in sufficiently complicated holographic models.

\section*{Acknowledgements}
 It is a pleasure for DM to acknowledge useful discussions with David Berenstein, Nathan Craig, William Donnelly, Tom Hartman, Rob Myers, Joe Polchinski, Vladimir Rosenhaus, Jorge Santos, Misha Smolkin, and Mark Srednicki.  He also indebted the participants of the Quantum Information in Quantum Gravity Workshop (Vancouver, BC, August 2014) for valuable feedback. Finally, we thank Aron Wall for editorial assistance and for many discussions on related subjects.  This work was supported in part by the National Science Foundation under grant numbers PHY12-05500 and PHY15-04541 and by funds from the University of California.  In addition, WB and ZF were funded respectively by the Caltech Summer Undergraduate Research Fellowship Program and the Tsinghua Xuetang Talents Program.  WB and ZF both thank the UCSB Physics department for its hospitality during the bulk of this work in summer 2014.

\providecommand{\href}[2]{#2}\begingroup\raggedright

\end{document}